\documentclass[twocolumn,preprintnumbers,amsmath,amssymb]{revtex4}

\usepackage{graphicx}
\usepackage{dcolumn}
\usepackage{bm}

\begin{document}

\preprint{\small{Accepted for publication in Applied Physics
Letters}}

\title{Organic small molecule field-effect transistors with Cytop$^{\mathrm{TM}}$ gate dielectric: eliminating gate bias stress effects}

\author{W. L. Kalb}%
 \email{kalb@phys.ethz.ch}
\author{T. Mathis}
\author{S. Haas}
\author{A. F. Stassen}
\author{B. Batlogg}
\affiliation{%
Laboratory for Solid State Physics, ETH Zurich, 8093 Zurich,
Switzerland
}%


\begin{abstract}
We report on organic field-effect transistors with unprecedented
resistance against gate bias stress. The single crystal and
thin-film transistors employ the organic gate dielectric
Cytop$^{\mathrm{TM}}$. This fluoropolymer is highly water repellent
and shows a remarkable electrical breakdown strength. The single
crystal transistors are consistently of very high electrical
quality: near zero onset, very steep subthreshold swing (average:
1.3\, nF\,V/(dec\,cm$^{2}$)) and negligible current hysteresis.
Furthermore, extended gate bias stress only leads to marginal
changes in the transfer characteristics. It appears that there is no
conceptual limitation for the stability of organic semiconductors in
contrast to hydrogenated amorphous silicon.
\end{abstract}

\maketitle


Field-effect mobilities in organic thin-films as high as in
hydrogenated amorphous silicon (a-Si:H) have been achieved for some
ten years now. The main advantage of organic semiconductors is the
easy deposition by thermal evaporation or printing. Early works
employed Si/SiO$_{2}$ substrates for convenience, and the device
characteristics significantly improved after Lin \textit{et al.}
suggested to render the SiO$_{2}$ surface hydrophobic with the
self-assembling agent octadecyltriclorosilane (OTS).\cite{LinYY1997}
To fully exploit the potential of organic semiconductors, however,
it is of great importance to employ easily processable (organic)
gate insulators. Thus, the search for suitable organic dielectrics
has intensified. Refs. \onlinecite{VeresJ2004} and
\onlinecite{YoonMH2006} are recent summaries.

One of the last obstacles to be overcome for a commercialization of
organic thin-film transistors is gate bias stress effects. Switching
the devices on for some time leads to a reduction in current at a
given gate voltage. Gate bias stress effects can result in a
significant difference between the forward and reverse
sweep\cite{HorowitzG1999, PesaventoPV2004, GoldmannC20062} and have
often been studied by applying a fixed gate voltage for an extended
time, followed by a measurement of the shift of the transfer
characteristic.\cite{VoelkelAR2002} The causes of gate bias stress
effects are not yet completely identified. The effects are thought
to be due to the trapping and release of charge carriers on a time
scale comparable to the measurement time. Mounting evidence
indicates that water in the dielectric-semiconductor interface
region can cause gate bias stress effects.\cite{GomesHL2006,
GoldmannC2006, PernstichKP2006}

In this letter we report on combinations of small molecule organic
semiconductors and an organic spin-on dielectric that yield
field-effect transistors with exceptionally high quality
characteristics and stability. The transistors have a bottom gate
structure with an amorphous fluoropolymer (Cytop$\mathrm{^{TM}}$) as
gate dielectric. This fluoropolymer is highly transparent, in
contrast to ordinary teflon, and its relative permittivity is
$\epsilon_{i}=2.1-2.2$.\cite{cytop} Cytop$^{\mathrm{TM}}$ has
previously been used in field-effect transistors with a polymeric
semiconductor in a top gate structure.\cite{VeresJ2004} We
demonstrate this favorable material in combination with two small
molecule semiconductors: rubrene and pentacene. The device stability
was evaluated by applying a gate bias for extended periods of time.


The devices were fabricated as follows: ITO coated glass slides
served as substrate and gate electrode. For the insulating layer,
Cytop CTL-809M (solvent: CT-Solv.180) from Asahi Glass, Japan was
spin-coated onto the ITO and dried for one hour at $90^{\circ}$\,C.
The thickness of the insulating layer was determined for each sample
with a surface step profiler. The films are 430 to 700\,nm thick,
which gives a gate capacitance of $C_{i}=4.4$ to $2.7$\,nF/cm$^{2}$.
Leakage current measurements on a typical sample ($457\pm10$\,nm
thick) show current levels below 1\,$\mu$A up to 450 V, above which
the dielectric breaks down.\cite{footnote1} 450\,V correspond to an
applied field of 9.8\,MV/cm and the current density at this field is
$2.7\times10^{-6}$\,A/cm$^{2}$. This is remarkably good for an
organic insulator and is better than the thermally grown SiO$_{2}$
that we generally use. The RMS roughness of the insulator was
investigated by AFM and is $\sim0.6$\,nm. Water contact angles are
between 109$^{\circ}$ and 116$^{\circ}$ (average: 112$^{\circ}$).

Rubrene and pentacene single crystal field-effect transistors
(SC-FETs) were made by evaporating 30\,nm thick gold source and
drain contacts onto the fluoropolymer in high vacuum. The single
crystals were grown separately by physical vapor transport with
argon as carrier gas.\cite{LaudiseRA1998} The crystals were placed
on the prefabricated substrates in air. Pentacene thin-film
transistors (TFTs) were made by evaporating a 50\,nm thick pentacene
film through a shadow mask onto the Cytop$^{\mathrm{TM}}$ in high
vacuum (base pressure $10^{-8}$\,mbar). The thin-film devices were
completed by evaporating gold electrodes. All transistor
measurements were carried out in a He atmosphere ($\mathrm{H_{2}O}$,
$\mathrm{O_{2}} <0.5$\,ppm) with a HP 4155A semiconductor parameter
analyzer. The step width was 0.5\,V and the integration time was
20\,ms (medium) with a zero delay time between the voltage steps.
The measurement of a transfer characteristic (forward and reverse
sweep) took $\sim70$\,s.

\begin{figure}
\includegraphics{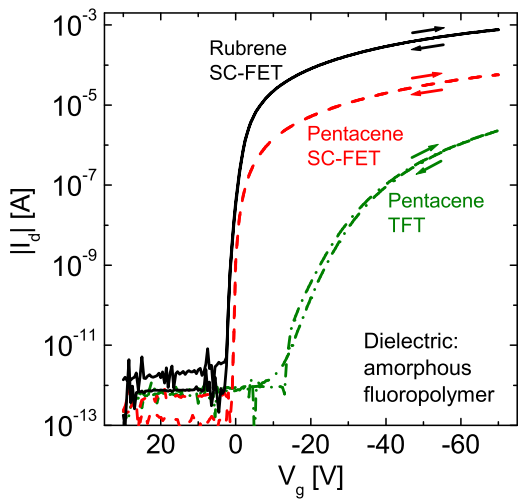}
\caption{\label{figure1} Transfer characteristics in saturation
($V_{d}=-80$\,V) for a rubrene SC-FET, a pentacene SC-FET and a
pentacene TFT. Forward and reverse sweeps are shown but are
indistinguishable over the entire operating range for the SC-FETs.
The TFT shows only a small current hysteresis near the onset.}
\end{figure}

\begin{figure}
\includegraphics{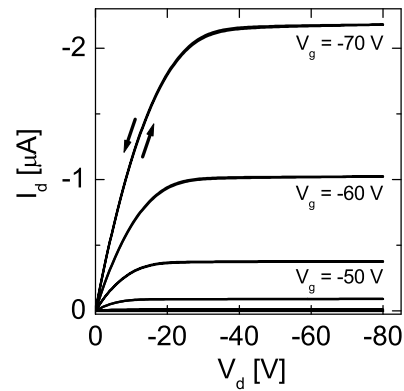}
\caption{\label{figure2} Output characteristic of the pentacene TFT
revealing the ideal thin-film transistor operation.}
\end{figure}

The excellent performance of the devices is shown in
Fig.~\ref{figure1}. The transfer characteristics from a rubrene
SC-FET, a pentacene SC-FET and a pentacene TFT, measured in
saturation with $V_{d}=-80$\,V, are given for the forward and the
reverse sweep. Most remarkable is the absence of any hysteresis for
the SC-FETs. We emphasize, that no additional steps such as
electrical aging or pre-stressing were taken to obtain these curves.
A further mark of the high quality of the devices is the steep
subthreshold swing, $S=0.50$\,V/dec for the rubrene SC-FET and
0.29\,V/dec for pentacene SC-FET. This gives normalized subthreshold
swings $C_{i}S$ of respectively 1.6\,nF\,V/(dec\,cm$^{2}$) and 1.3\,
nF\,V/(dec\,cm$^{2}$). Higher subthreshold swings and a large
current hysteresis are measured when we place nominally identical
crystals on OTS surface-treated SiO$_{2}$.\cite{GoldmannC2006,
GoldmannC20062} Remarkable as well is the very small (slightly
positive) onset voltage of the two single crystal devices (+3\,V for
rubrene and +1\,V for pentacene). The saturation field-effect
mobilities from the crystals ($5.7$\,cm$^{2}$/Vs and
$1.4$\,cm$^{2}$/Vs) are comparable to pentacene or rubrene SC-FETs
with other dielectrics. The thin-film transistor shows a notable
hysteresis in the transfer characteristic close to the onset
voltage, and the onset is more negative than in the case of the
single crystals, i.e. $-13$\,V (Fig.~\ref{figure1}). The hysteresis
is much less apparent on a linear scale, i.e when the transistor is
switched on completely. Fig.~\ref{figure2} shows the output
characteristic of the pentacene TFT, revealing the ideal thin-film
transistor behavior, with a saturation field-effect mobility of
$0.26$\,cm$^{2}$/Vs.\cite{footnote2}

The linear regime transfer characteristics measured with
$V_{d}=-5$\,V (not shown) yield almost identical results in terms of
subthreshold swing and current hysteresis. Only for the pentacene
SC-FET we observe a slight increase in hysteresis in the on-current.
In the subthreshold regime however, the hysteresis is not increased.

We made 17 rubrene SC-FETs on 7 substrates with Cytop films, and the
crystals were grown in 4 different runs. All transistors, except for
a pathologic one, have a steep subthreshold swing, a near zero onset
voltage and a very small hysteresis. The normalized subthreshold
swings from the 16 devices are between 0.75 and 2.6\,
nF\,V/(dec\,cm$^{2}$) with an average value of 1.3\,
nF\,V/(dec\,cm$^{2}$). These values are among the best values for an
organic transistor obtained to date.\cite{PodzorovV2003,
BrisenoA2006} In some cases, there is a slightly increased
hysteresis in the on-current.

The advantages of the material combinations become striking in gate
bias stress studies. We have applied a gate voltage to the three
devices in Fig.~\ref{figure1} for a prolonged time. After the
initial transfer characteristic measurement, a gate bias of $V_{g} =
-70$\,V was applied for two hours. After a two hour relaxation
period, a gate bias of $V_{g} = +70$\,V was applied. During the
stress periods, the source was grounded and the drain potential was
held at 0\,V to ensure homogenous gate stress. The drain current
depends (approximately) quadratically on the effective gate voltage
and is very sensitive to changes induced by a two hour stress
period. Bias stress experiments were carried out in the
dark.\cite{footnote3}

\begin{figure}
\includegraphics{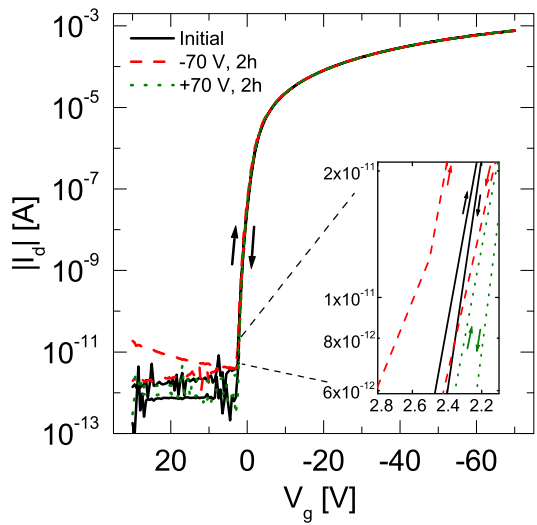}
\caption{\label{figure3} The rubrene single crystal device is highly
stable against gate bias stress. The main panel shows the transfer
characteristic measured at $V_{d}=-80$\,V prior to the stress
sequence (full black line), after two hours of gate bias stress at
$V_{g}=-70$\,V (dashed red line) and after subsequent gate bias
stress at $V_{g}=+70$\,V for two hours (dotted green line). The
graph includes the forward and reverse sweep in all three cases. The
inset shows the drain currents close to the onset voltage.}
\end{figure}

For the rubrene SC-FET, Fig.~\ref{figure3} shows the initial
characteristic, the characteristic measured after 2 hours of
negative bias and after 2 hours of positive bias. The device is
hardly influenced by the long application of a gate bias. There are
only marginal changes in the transfer characteristic. After negative
stress, there is a very small shift of the onset voltage to more
positive voltages, accompanied by a small increase in current
hysteresis and a small decrease in on-current. For the pentacene
SC-FET, the observations are similar. When compared to the rubrene
device, the shift of the onset voltage due to bias stress is even
smaller but the decrease in on-current is somewhat more pronounced
(3.8\,\% at $V_{g}=-70$\,V). In contrast, in similar experiments
with single crystals of rubrene or pentacene on OTS-treated
SiO$_{2}$, large shifts of the transfer characteristics are
observed.\cite{GoldmannC20062, GoldmannC2006} For the pentacene TFT,
a gate voltage of $V_{g} = -70$\,V applied for two hours leads to a
rigid shift of the curve by $-5.2$\,V to more negative voltages.


The surface of the amorphous fluoropolymer proves to have a highly
desirable quality: essentially no electrically active trap states
form in combination with the organic semiconductors. Bias stress
effects in SC-FETs are marginal, and thus long-lived states for
holes are (almost) non-existent at the insulator surface. The
absence of insulator surface states can also account for an improved
subthreshold swing. It is remarkable that the insulator works very
well with two different semiconductors, i.e. rubrene and pentacene.
This may indicate that the absence of surface states is due to the
absence (or low density) of a specific chemical species on the
insulator surface. Gate bias stress effects in SC-FETs are known to
be less severe when the hydrophilic SiO$_{2}$ is rendered
hydrophobic with OTS (water contact angle of
$90^{\circ}-95^{\circ}$).\cite{GoldmannC20062, GoldmannC2006} The
highly hydrophobic Cytop$^{\mathrm{TM}}$ surface (water contact
angle of $\sim112^{\circ}$) leads to an (almost) complete
elimination of gate bias stress effects. Water is a known cause of
gate bias stress effects which can be eliminated by employing a
highly hydrophobic Cytop$^{\mathrm{TM}}$ gate insulator. The high
reproducibility of the excellent device performance matches the
reproducibly high water contact angle of the Cytop$^{\mathrm{TM}}$
films.

For the pentacene SC-FET and TFT we have combined the same
semiconductor and insulator. However, we observe that the TFT is
less stable against bias stress than the SC-FET. Also the onset
voltage of the TFT is more negative. These effects cannot be
attributed to insulator surface states but should be due to
localized states within the pentacene layer close to the
dielectric-semiconductor interface.


In conclusion, this study highlights the generically high
performance and high stability of small molecule organic
semiconductors when combined with a suitable gate dielectric. FETs
with rubrene and pentacene in combination with a fluoropolymer show
excellent electrical characteristics, and they are hardly affected
by long-term gate bias stress. It seems that there is no conceptual
limitation for the stability of organic semiconductors in contrast
to a-Si:H, where the diffusion of hydrogen leads to gate
bias-induced metastable defects.\cite{JacksonWB1989}


The authors thank K. P. Pernstich for help with the experimental
setup, K. Mattenberger and H. P. Staub for support in technical
issues and Asahi Glass for providing the Cytop$^{\mathrm{TM}}$
samples.


\begin{thebibliography}{18}
\bibitem[{foo({\natexlab{b}})}]{footnote2}
\bibinfo{note}{The TFTs were transported through air but were kept in the dry
  He atmosphere for several days prior to electrical characterization. The
  storage led to a slight improvement in hysteresis.}
\bibitem[{foo({\natexlab{c}})}]{footnote3}
\bibinfo{note}{Both the negative and the positive bias stress period were
  shortly interrupted ($70$\,s) after 10, 20, 40, and 60 minutes in order to
  measure a transfer characteristic.}
\end{thebibliography}






\end{document}